\documentclass[a4paper,10pt,twoside]{cpc-hepnp}
\usepackage{CJK,upgreek,fancyhdr}
\usepackage{subfigure}
\usepackage{multicol}
\usepackage{booktabs}
\usepackage{amsmath}
\usepackage{amsfonts,amssymb,bm,mathrsfs,bbm,amscd}
\usepackage{lastpage}
\usepackage[pdftex]{graphicx}
\usepackage{multirow}
\DeclareGraphicsExtensions{.pdf,.jpeg,.png}
\usepackage{epstopdf}
\usepackage[english]{babel}
\usepackage[colorlinks=false,pdfpagemode=FullScreen,,setpagesize=off,pdfborder={0 0 0}]{hyperref}
\usepackage{overpic}
\usepackage{lineno}
\usepackage{color}
\usepackage{diagbox}
\usepackage{makecell}

\begin{document}
\begin{CJK*}{UTF8}{gkai}


\title{Identification of deuteron at BESIII} 

\maketitle

\begin{center}

\author{
	\begin{small}
		\begin{center}
			Wei Wang$^{1}$(王维),
			Bo Zheng$^{1}$$^{,}$\footnotemark[1](郑波),
			Jia Jia Qin$^{1}$(秦佳佳),
			Zhi Yong Wang$^{2,3}$(王至勇)\\
				\vspace{0.2cm} {\it
					$^{1}$ University of South China, Hengyang 421001, People's Republic of China\\
					$^{2}$ Institute of High Energy Physics, Beijing 100049, People's Republic of China\\
					$^{3}$ University of Chinese Academy of Sciences, Beijing 100049, People's Republic of China\\
				}
			\end{center}
			\vspace{0.4cm}
		\end{small}
	}
	\renewcommand{\thefootnote}{\fnsymbol{footnote}}
	\footnotetext[1]{Corresponding author(1):zhengbo\_usc@163.com}
\end{center}

\begin{abstract}
The identification of deuteron with momentum in range of $0.52$-$0.72$ GeV/$c$ has been studied with the specific ionization energy loss information, using the data sample collected by the BESIII detector at center-of-mass energies between $4.009$ and $4.946$ GeV. Clean deuteron samples are selected using time of flight information. For all data samples, the deuteron identification efficiencies are higher than $95\%$, with a maximum difference of $(4.93\pm0.95)\%$ between data and MC simulation. These verify the effectiveness of the deuteron identification method based on the specific ionization energy loss, and provide valuable information for the future studies of the processes involving deuteron in the final state at BESIII.
\end{abstract}


\begin{multicols}{2}

\section{Introduction}\label{sec.I}
The deuteron ($d$) is a bound state of proton and neutron, and has long been a prominent subject of physics. In the field of nuclear physics, the coalescence model was initially proposed to explain the emission of $d$ in reactions induced by protons with energies of $25$-$30$ GeV~\cite{yy1,yy2}. Over the years, the coalescence model has become an effective tool for describing the production of particles and anti-particles~\cite{yy3}. The study of $d$ production is crucial to verify theoretical models~\cite{yy4} and provides a basis for understanding the particle production mechanism~\cite{yy5}. In the fields of astrophysics and cosmology, the study of the processes involving cosmic anti-nuclei, such as anti-deuteron ($\bar{d}$), serves as sensitive probes for dark matter annihilation and allows for indirect studies of dark matter~\cite{yy6,yy7}. This kind of study is also important to comprehend the properties of dense astrophysical objects, such as neutron stars~\cite{yy8,yy9}. In the field of particle physics, experimental studies of the processes involving $d$ are critical to validate the Lund string fragmentation model and predict the $d$ production in $Z$ boson decays.~\cite{yy10}. Many results have been reported by various experiments, including heavy ion collision~\cite{yy11}, proton-proton collision~\cite{yy12}, proton-nucleus collision~\cite{yy13}, and photoproduction reaction~\cite{yy14}. In contrast, the relevant study in positron-electron collision is relatively limited, due to its low production cross-section~\cite{yy15}.

In recent years, more and more hadrons, such as $q\bar{q}$ for mesons, $qqq$ for baryons, and $q\bar{q}q\bar{q}$ for ``tetraquark” states~\cite{yy17}, $qqqq\bar{q}$ for ``pentaquark” states~\cite{yy18}, have been observed in experiments. Especially, the exotic state $d$*($2380$), with mass around $2380$ MeV/$c^{2}$ and width of about $70$ MeV, was observed in the isoscalar double-pionic fusion process $pn \rightarrow d \pi^{0} \pi^{0}$~\cite{yy19} and was subsequently confirmed in many other processes~\cite{yy20,yy21,yy22,yy23}. This state has been proposed as an excited $d$, a molecule with large $\Delta\Delta$ component~\cite{yy24}, or a hexaquark state dominated by hidden-color component~\cite{yy25}. Because one-third of $d$*($2380$) decays into final states involving $d$, the processes involving $d$ offers potential test-bed to investigate the properties of $d$*($2380$). The BESIII experiment, operated in the tau-charm energy region, has collected large positron-electron collision data sample in the $4.009$-$4.946$ GeV energy range. This provides good opportunity to study the $d$ and $\bar{d}$ productions, the $d$*($2380$) resonance, the dibaryon states, and the hexaquark states.

Good performance of particle identification (PID) is essential for the precision measurements in quark flavour physics, $\tau$ physics, top physics, Higgs physics, and other fields~\cite{yy26}. The BES and BESIII Collaborations reported systematic studies of the PID efficiencies of electron, muon, pion, kaon and proton~\cite{yy27,yy28,yy29,yy30}. The LHCb and BaBar Collaborations have also reported extensive PID studies~\cite{yy31,yy32}. However, little knowledge about the $d$ PID is currently available due to low production rate. An effective $d$ identification method is helpful to reduce the backgrounds from other particles. Previously, the ARGUS~\cite{yy33}, BaBar~\cite{yy34}, CLEO~\cite{yy35} and ALEPH~\cite{yy36} Collaborations reported the $\bar d$ productions, while only the ALEPH Collaboration reported the $d$ PID study. Usually, the specific ionization energy loss ($dE/dx$) and time-of-flight (TOF) measurements were used for $d$ or $\bar d$ identification. Details of methods and momentum ranges in the $d$ or $\bar d$ PID studies from different experiments are shown in Table \ref{TABYI}.

In this paper, we study the $d$ PID efficiencies with the $dE/dx$ method, using the data sample collected by the BESIII detector at center-of-mass (c.m.) energies between $4.009$ and $4.946$ GeV. The $\bar d$ is not considered due to limited statistics. In addition, due to the challenges of detecting low-momentum particles by the TOF detector and the limited statistics of high-momentum $d$, we only aim for the PID efficiencies of $d$ with momentum ranging from 0.52 to 0.72 GeV/$c$.
\begin{center}
\centering
\footnotesize
\tabcaption{\label{TABYI} Methods and momentum ranges of the $d$ or $\bar{d}$ PID studies from different experiments.}
\begin{tabular}{c|c|c|c}
\hline
\hline
Experiment&Method&Object&$p$~(GeV/$c$)\\
\hline
ARGUS&$dE/dx$+TOF&$\bar{d}$&$0.45$-$1.70$\\
\hline
BaBar&$dE/dx$&$\bar{d}$&$0.50$-$1.50$\\
\hline
CLEO&$dE/dx$&$\bar{d}$&$0.45$-$1.45$\\
\hline
ALEPH&$dE/dx$&$d$/$\bar{d}$&$0.62$-$1.03$\\
\hline
\hline
\end{tabular}
\end{center}

\section{The BESIII detector}\label{sec.II}

The Beijing Spectrometer III (BESIII)~\cite{yy37} is a general-purpose detector operated at the Beijing Electron-Positron Collider II~\cite{yy38}. The BESIII detector consists of four sub-detectors: the mult-layer drift chamber (MDC), the TOF counter, the electric-magnetic calorimeter, and the muon chamber. The MDC sub-detector determines the momentum, and vertex position for charged particles~\cite{yy39}. It also provides $dE/dx$ information for PID of charged particles. It is a type of gas detector that contains multiple layers of field wires and signal wires. Its operating gas consists of a mixture of helium (He) and C$_{3}$H$_{8}$ in a ratio of $60$:$40$. The energy loss of charged particles through ionization in the working gas, $dE/dx$, is obtained by measuring the charge deposited on the signal wires. After $dE/dx$ calibration, a resolution of about $6\%$ has been obtained for minimum ionization particles. The TOF counter measures the flight time of charged particles, which is widely used for PID. The time resolution in the TOF barrel~\cite{yy37} region is $68$ ps, whereas that in the end-cap~\cite{yy37} region is $110$ ps. In $2015$, the end-cap TOF system underwent an upgrade utilizing multi-gap resistive plate chamber technology, resulting in an improved time resolution of $65$ ps~\cite{yy40}.

The BESIII detector is simulated by the GEANT4-based simulation software BOOST~\cite{yy41, yy42}, which includes the geometric and material description of the BESIII detector, the detector response and digitization models.

\section{Data samples and simulation}\label{sec.III}
The data samples taken at c.m. energies between $4.009$ and $4.946$ GeV with a total integrated luminosity about $18$ $\rm fb^{-1}$ are used in this analysis. For data sample at each energy point, the c.m. energy is measured using $e^+e^-\rightarrow\mu^+\mu^-$ process, with an uncertainty less than $1.0$ MeV~\cite{yy43,yy44}, and the integrated luminosity is measured using Bhabha process, with an uncertainty of $1.0\%$~\cite{yy44}. The data samples are divided into seven subsamples according to different data taking periods, to consider slightly different detector performence.

The $e^+e^-\rightarrow{\bar p}{\bar n}d$~process is simulated in phase-space model with ConExc~\cite{yy45,yy46} at each c.m. energy. The potential backgrounds are studied using the inclusive Monte Carlo (MC) sample, corresponding to an integrated luminosity of 32 fb$^{-1}$ at $\sqrt s=4.178$ GeV with KKMC~\cite{yy47,yy48}.

\section{Identification of deuteron}
\subsection{Identification method}
The BESIII experiment usually combines both $dE/dx$ and TOF information to identify charged particles. As shown in Fig.~\ref{FIGYI}, both $dE/dx$ and TOF information can well separate $d$ from other particles in a certain momentum range. However, due to the relatively large mass, few deuterons reach the TOF detector. Identifying $d$ with the TOF information will cause significant efficiency loss. Therefore, we use the $dE/dx$ information to study the $d$ PID, with the help of the TOF information. See section \ref{PID efficiency} for more information.
\begin{center}
\centering
\includegraphics[width=0.23\textwidth]{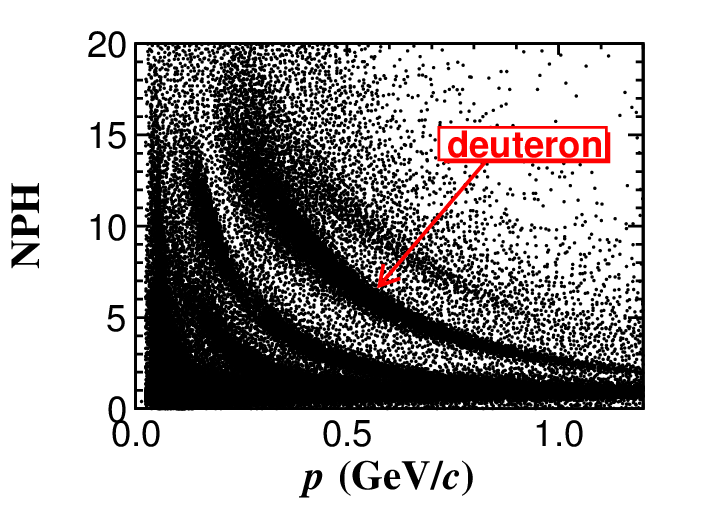}
\includegraphics[width=0.23\textwidth]{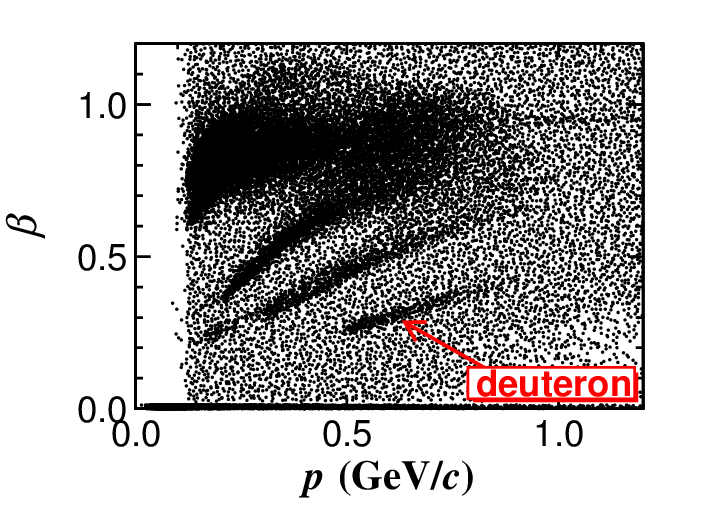}

\includegraphics[width=0.23\textwidth]{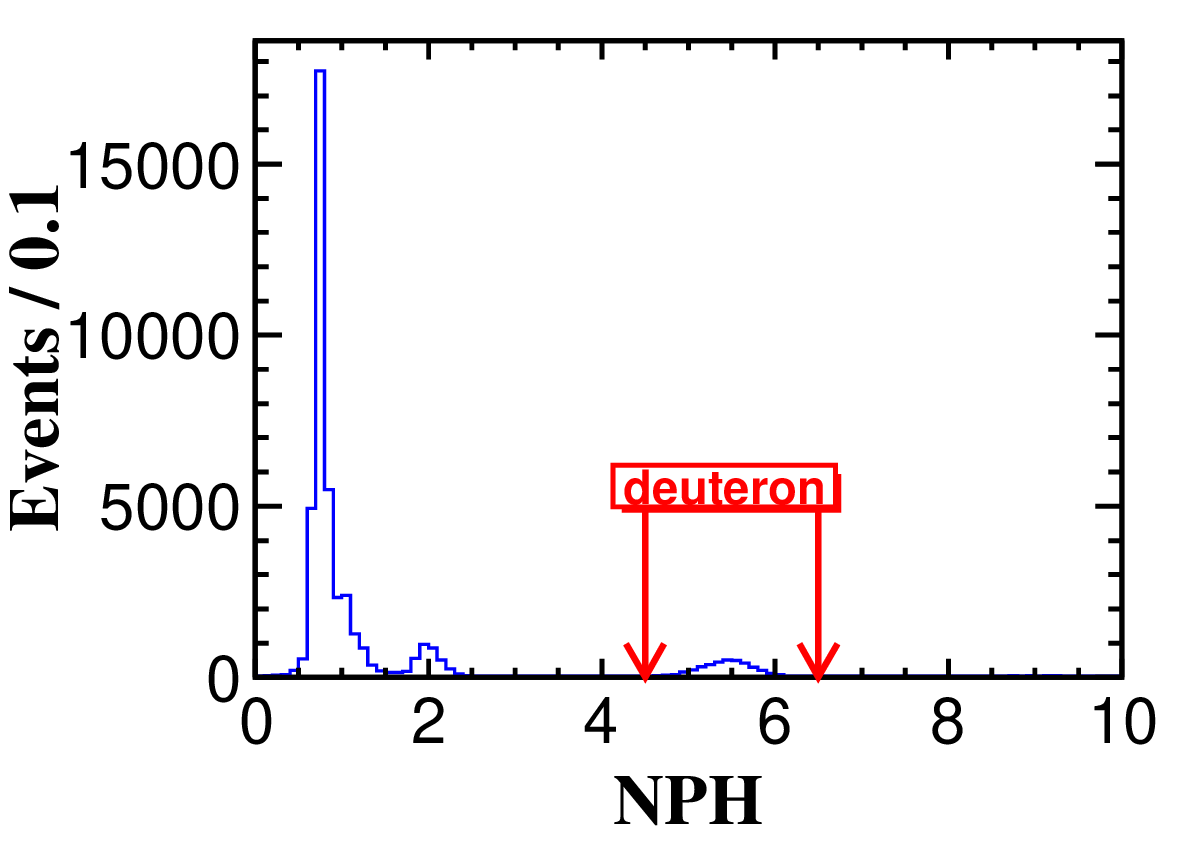}
\includegraphics[width=0.23\textwidth]{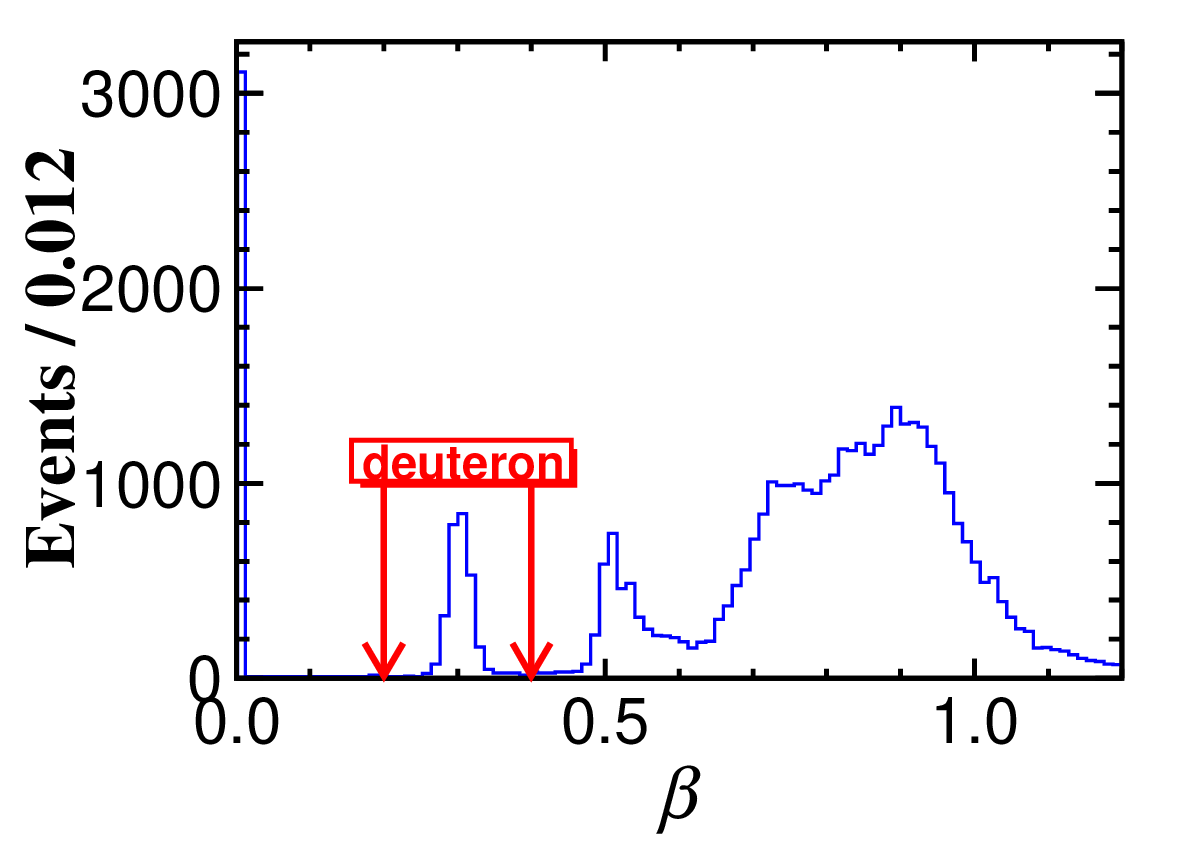}
\figcaption{The distributions of (top left) momentum versus normalized pulse height of $dE/dx$ (NPH) and (top right) momentum versus $\beta$. The distributions of (bottom left) NPH and (bottom right) $\beta$ for candidates in the momentum range of 0.595-0.605 GeV/$c$. All plots are based on the 2016 data sample.}
\label{FIGYI}
\end{center}

The $dE/dx$ method uses the normalized residual of $dE/dx$, denoted as
\begin{equation}
\chi_{dE/dx} = \dfrac{dE/dx_{\rm meas} - dE/dx_{\rm exp}}{\sigma_{dE/dx}},
\label{EQYI}
\end{equation}
where $dE/dx_{\rm meas}$, $\sigma_{dE/dx}$ and $dE/dx_{\rm exp}$ represent the measured value, the uncertainty of $dE/dx_{\rm meas}$ and the expected value, respectively. The $\chi_{dE/dx}$ distribution is expected to follow a normal distribution, as shown in Fig.~\ref{FIGYIV}.

\begin{center}
\centering
\includegraphics[width=0.23\textwidth]{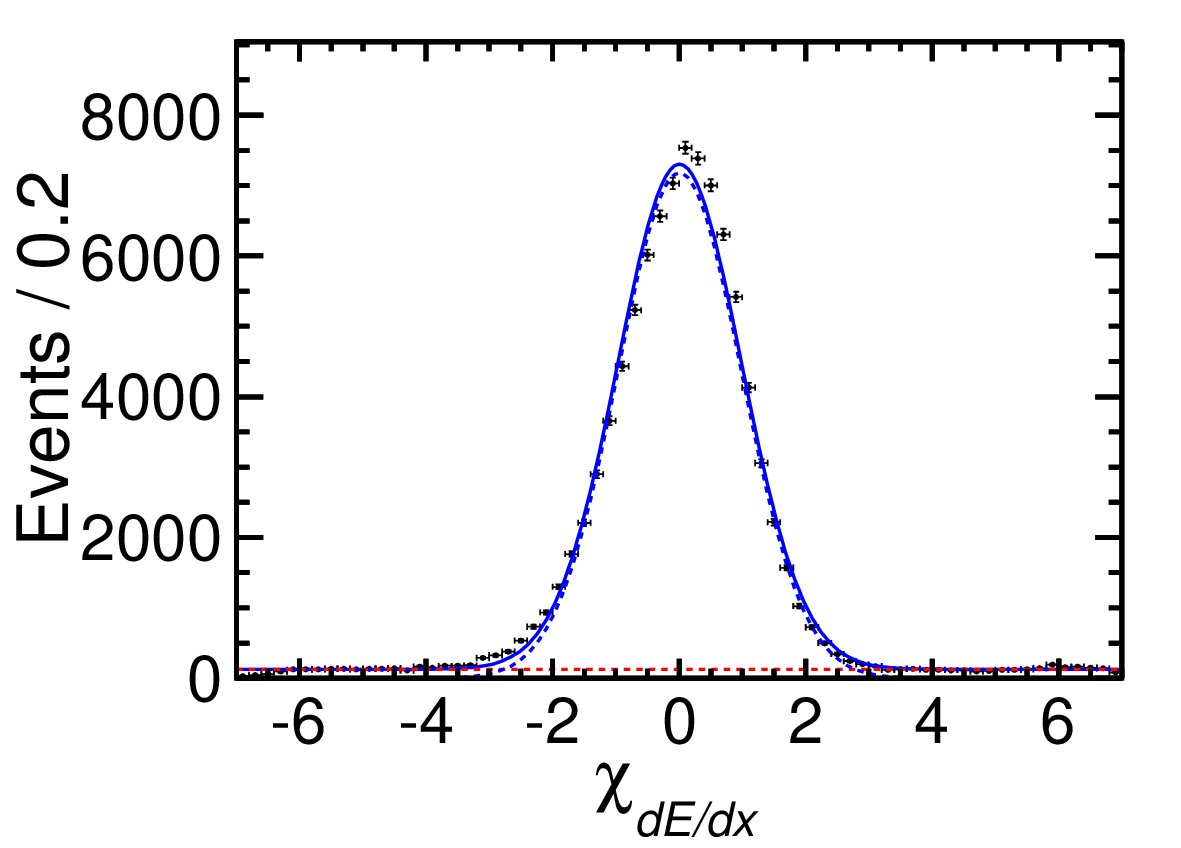}
\includegraphics[width=0.23\textwidth]{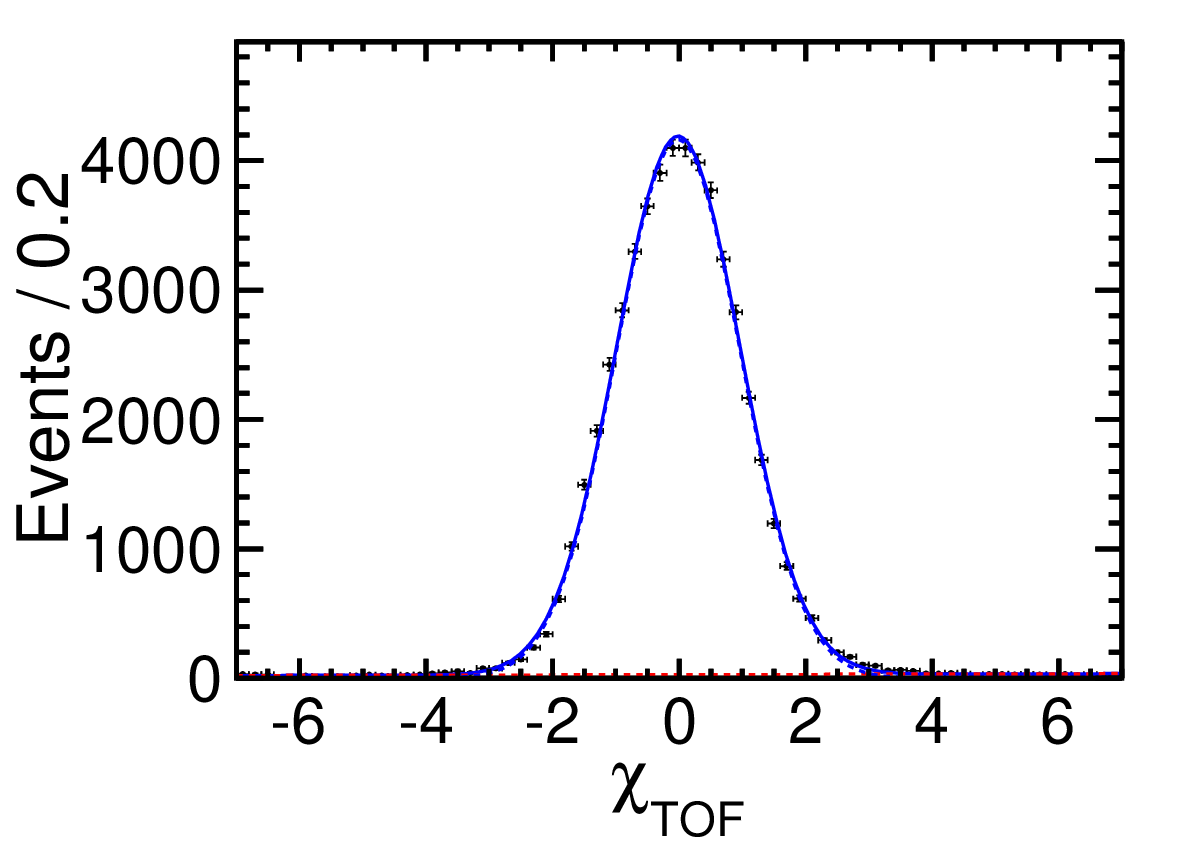}
\figcaption{Fits to $\chi_{dE/dx}$ (left) and $\chi_{\rm TOF}$ (right) of the $2016$ data sample (dots with error bars). A Gaussian function marked with blue dashed line and a first-order Chebychev polynomial marked with red dashed line.}
\label{FIGYIV}
\end{center}

Once the $\chi_{dE/dx}$ distribution is obtained, the probability density function of a given particle hypothesis is constructed as
\begin{equation}
{\rm prob}(\chi_{dE/dx}^{2}, 1) = 1 - f(\dfrac{1}{2}, x),
\label{EQGG}
\end{equation}
with $x=\chi_{dE/dx}^{2}/2$ and $f(\dfrac{1}{2},x) = \dfrac{1}{\sqrt{\pi}}\times\int_{0}^{x}t^{-\dfrac{1}{2}}\times{e^{-t}dt}$. The `prob' represents the probability of a value greater than the observed $\chi_{dE/dx}^{2}$. It is required to satisfy a certain value to achieve the desired identification ability. In the $d$ identification, the ${\rm prob}(\chi_{dE/dx}^{2}, 1)$ is usually required to be greater than a certain value in the range of $(0, 1)$, and a larger value represents a more stringent PID requirement.

The deuterons are expected to be primarily produced through the reaction of charged particles with the beam pipe materials. As a result, the flight time is divided into two segments. However, since the flight distance in the beam pipe is extremely short, the impact on the time resolution is negligible. Therefore, this effect is disregarded in this study. As shown in Fig.~\ref{FIGYI}, the $\beta$ represents the ratio of the speed of charged particles to the speed of light, defined as
\begin{equation}
 \beta = \dfrac{L_{\rm path}}{c \times t_{\rm TOF}},
 \label{EQER}
\end{equation}
where $L_{\rm path}$, $c$ and $t_{\rm TOF} $ represent the flight distance of the particle, the speed of light and the flight time of the charged particle, respectively. The $\chi_{\rm TOF}$ value is given by
 \begin{equation}
\chi_{\rm TOF} = \dfrac{\beta_{\rm meas} - \beta_{\rm exp}}{\sigma_{\rm TOF}},
 \label{EQSA}
\end{equation}
where $\beta_{\rm meas}$, $\sigma_{\rm TOF}$ and $\beta_{\rm exp}$ represent the measured value, the uncertainty of $\beta_{\rm meas}$ and the expected value, respectively. Similarly, the $\chi_{\rm TOF}$ distribution is expected to follow a normal distribution, as shown in Fig.~\ref{FIGYIV}.

\subsection{Momentum correction}
When the charged particles pass through a detector, they interact with detector materials, thereby causing momentum loss. To account for this effect, BESIII has developed a track fitting algorithm based on the Kalman filter method~\cite{yy49}. This algorithm carefully handles the effects of multiple scattering, energy loss, non-uniformity of magnetic field and wire sag. In order to improve the momentum resolution and reduce the mean value of the difference between truth and reconstructed momenta in candidate events, we apply the track fitting algorithm for the $d$ reconstruction. The input/output check shows that this algorithm works well in the $d$ momentum correction. As shown in Fig.~\ref{FIGCOR}, the difference between the corrected and true momenta has been significantly improved in the simulated sample.
\begin{center}
\centering
\includegraphics[width=0.23\textwidth]{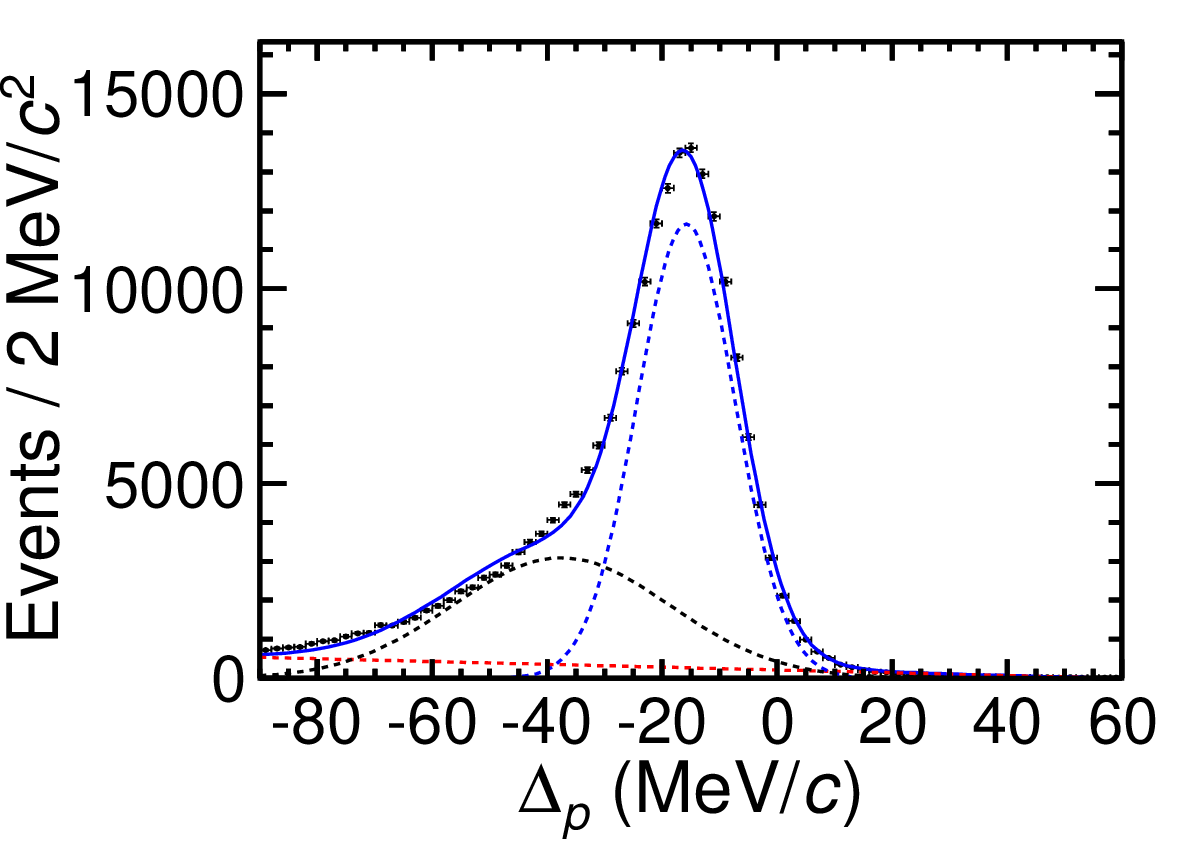}
\includegraphics[width=0.23\textwidth]{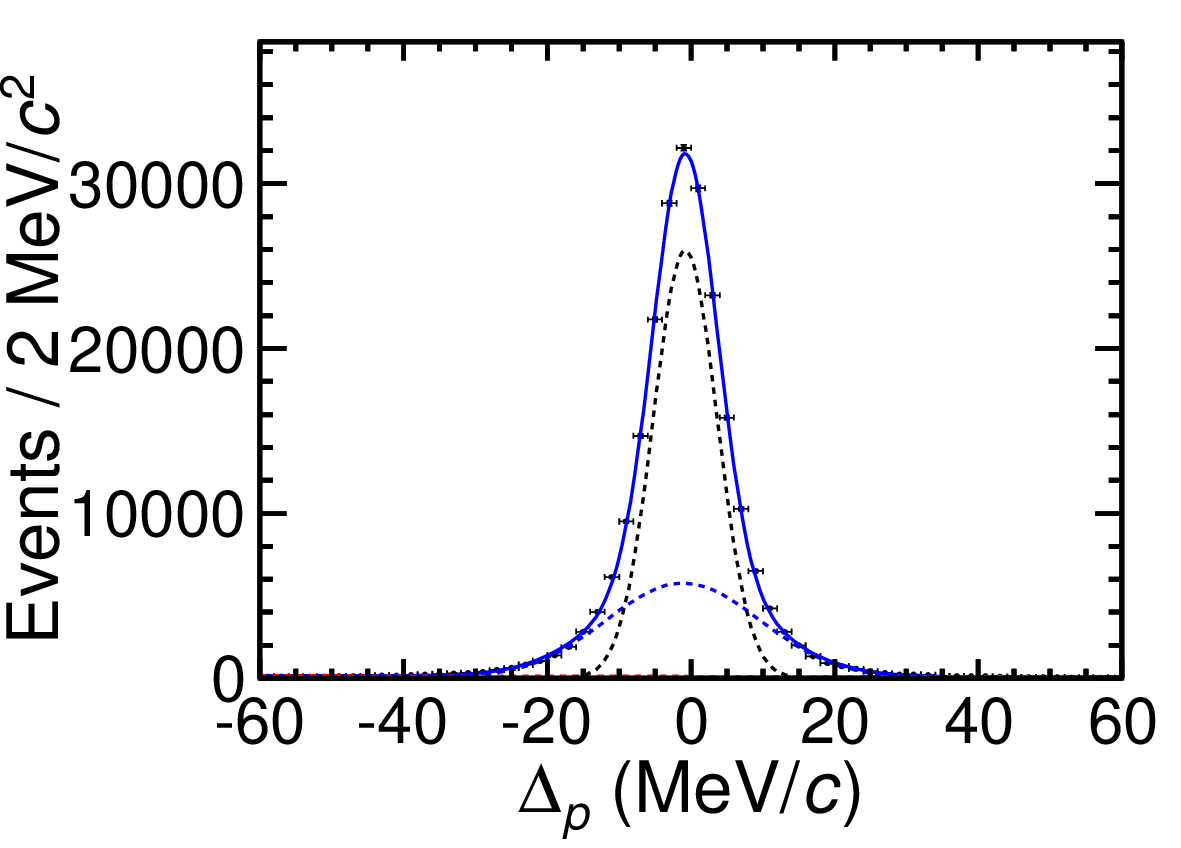}
\figcaption{Momentum difference between reconstruction and truth before (left) and after (right) correction. Two Gaussian functions marked by blue dashed and black dashed lines and a first-order Chebychev polynomial marked by red dashed line.}
\label{FIGCOR}
\end{center}

After the momentum correction using the track fitting algorithm, we study the PID efficiencies of $d$ in momentum range of $0.52$-$0.72$ GeV/$c$. 

\subsection{Event selection} \label{Event selection}
To select $d$ candidates, the number of charged tracks is required to be larger than two for data, while it is required to be two and the net charge is required to be zero for MC simulation. Charged tracks are reconstructed with the MDC hits within the range of $|\cos\theta|$ \textless $0.93$, where $\theta$ is the polar angle with respect to the $z$-axis. They are required to originate from the interaction region, defined as $R_{xy}$ \textless $1.0$ cm and $\vert V_z \vert$ \textless $10.0$ cm, where $R_{xy}$ and $\vert V_z \vert$ are the projections of the distances from the closest approach of the tracks to the interaction point in the $x$-$y$ plane and in the $z$ direction, respectively. The combined $dE/dx$ and TOF information is used to identify the observed particles, to improve the purity of the sample. The probabilities of identifying the track as electron, pion, kaon, and proton are required to be less than $0.001$, which ensures high purity of the selected $d$ samples.

After applying all above selection criteria, the remaining backgrounds are studied with the large inclusive MC sample generated at $\sqrt s=4.178$ GeV. In the $d$ signal region, the distribution of momentum versus NPH of the inclusive MC sample exhibits a distinct banded distribution. With the topology and event visualization tool~\cite{yy50}, we find that these tracks predominantly originate in the beam pipe. This is further supported by the two-dimensional distribution of $V_{x}$ versus $V_{y}$ in the data sample, and the $d$ sample is significantly larger than the $\bar d$ sample. Therefore, the deuterons used are mainly produced by the reaction of secondary charged particles with the beam pipe materials.

To improve the purity of the $d$ sample, some further selection criteria of $p_d\in (0.52,0.72)$~GeV/$c$, $\chi_{dE/dx}\in(-3,3)$, and $\chi_{\rm TOF}\in (-3,3)$ have been imposed on the selected candidates. The transverse momentum and cos$\theta$ distributions of the accepted candidates in data and MC simulation are shown in Fig.~\ref{FIGSJI}. The data-MC consistency in the $\cos\theta$ distribution is good, while the transverse momentum is distributed in the same interval with small difference in shape. The difference in the transverse momentum is mainly caused by the different sources of $d$ production in data and MC simulation. In data, the $d$ is not subject to channel restrictions and is mostly produced through secondary particle reactions with the beam pipe. In MC simulation, the $d$ is produced through exclusive process at the collision point. This discrepancy leads to small difference in the transverse momentum distributions between data and MC simulation.
\begin{center}
\centering
\includegraphics[width=0.23\textwidth]{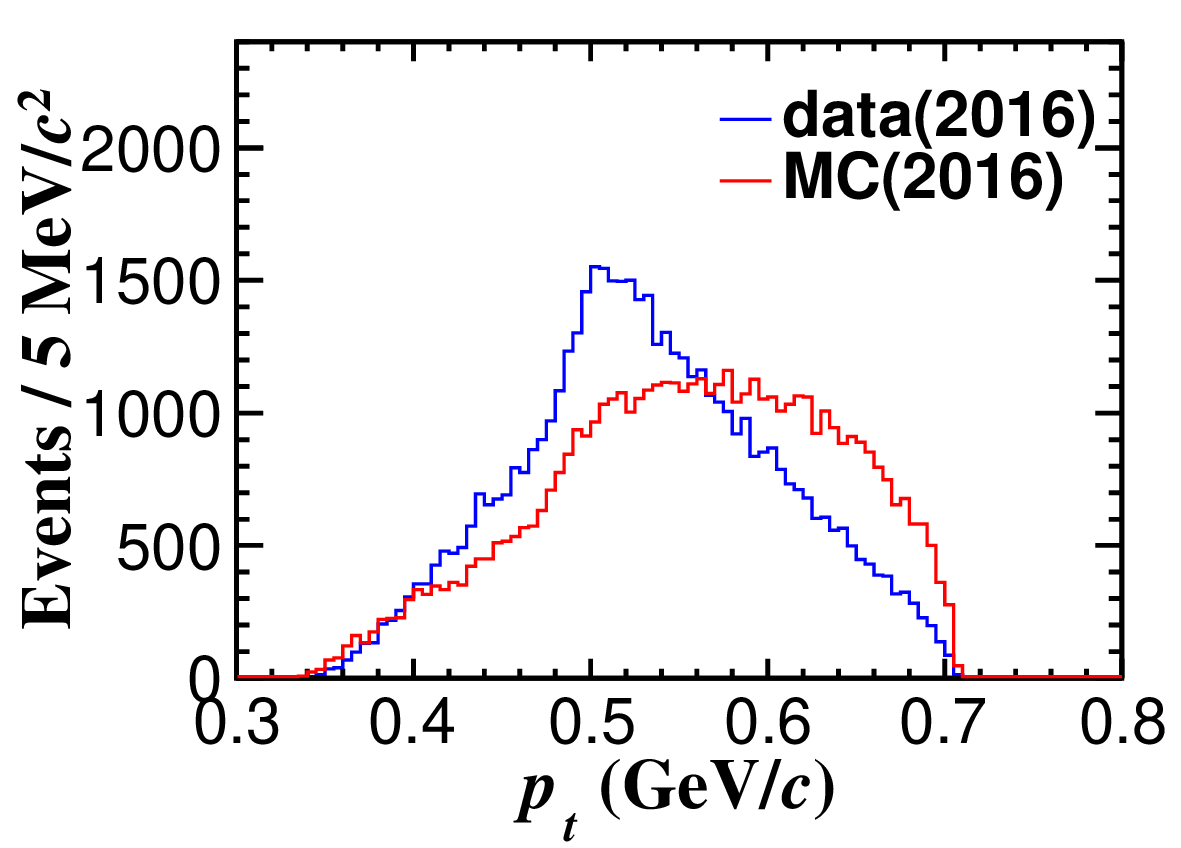}
\includegraphics[width=0.23\textwidth]{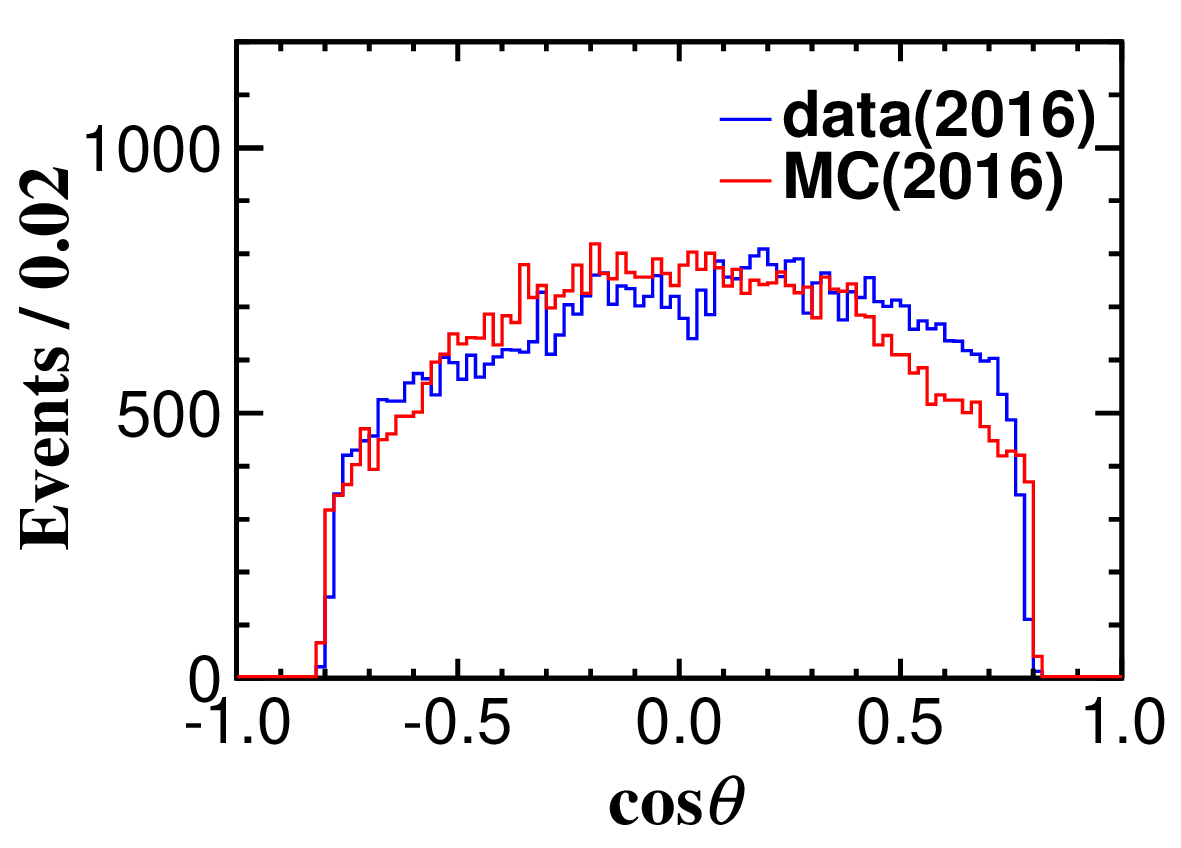}
\figcaption{The transverse momentum (left) and cos$\theta$ (right) distributions of candidates in the 2016 data sample and signal MC sample.}
\label{FIGSJI}
\end{center}

The purity of $d$ samples, as shown in Table \ref{TABER}, is obtained by integrating the corresponding $\chi$ distribution within $\pm$$3\sigma$ range. The purity of the obtained $d$ samples is higher than $97\%$.
\begin{center}
\centering
\normalsize
\tabcaption{\label{TABER} Purity of $d$ samples in different $\chi_{dE/dx}$ and $\chi_{\rm TOF}$ signal intervals.}
\begin{tabular}{c|c|c}
\hline
\hline
\multirow{2}{2em}{Year}&\multicolumn{2}{c}{Purity~($\%$)}\\
\cline {2-3}
&$dE/dx$&TOF\\
\hline
$2011$&$99.2$&$98.5$\\ 
\hline
$2013$&$98.6$&$98.6$\\ 
\hline
$2014$&$99.3$&$98.6$\\ 
\hline
$2016$&$99.4$&$98.4$\\ 
\hline
$2017$&$98.7$&$98.4$\\ 
\hline
$2020$&$97.7$&$98.4$\\ 
\hline
$2021$&$97.9$&$98.2$\\ 
\hline
\hline
\end{tabular}
\end{center}

\subsection{PID efficiency} \label{PID efficiency}
To calculate the $d$ PID efficiency using the $dE/dx$ information, we need to obtain the total number of deuterons at first. Since no exclusive control sample is available, the total number of deuterons can only be obtained from data. Considering that the $d$ PID method only with TOF information has been well established, the $d$ sample selected  by the TOF information is used as the control sample. The $d$ PID efficiency is determined by
\begin{equation}
\varepsilon = \dfrac{N_{\rm obs}}{N_{\rm tot}},
\label{EQSI}
\end{equation}
where $N_{\rm tot}$ is the number of signals  obtained from the fit to the $\chi_{\rm TOF}$ distribution, and $N_{\rm obs}$ is the number of signals obtained from the fit to the $\chi_{dE/dx}$ distribution of the accepted candidates. $N_{\rm tot}$ is obtained by integrating the fitted signal shape within $({\rm mean}-3\sigma, {\rm mean}+3\sigma)$, while $N_{\rm obs}$ is obtained from the fit to the obtained $\chi_{dE/dx}$ distribution. The ``mean" value corresponds to the expected value obtained from the Gaussian fit. As an example, Fig.~\ref{FIGESY} shows the fits to the $\chi_{dE/dx}$ and $\chi_{\rm TOF}$ distributions of the 2016 data sample. 
\begin{center}
\centering
\includegraphics[width=0.23\textwidth]{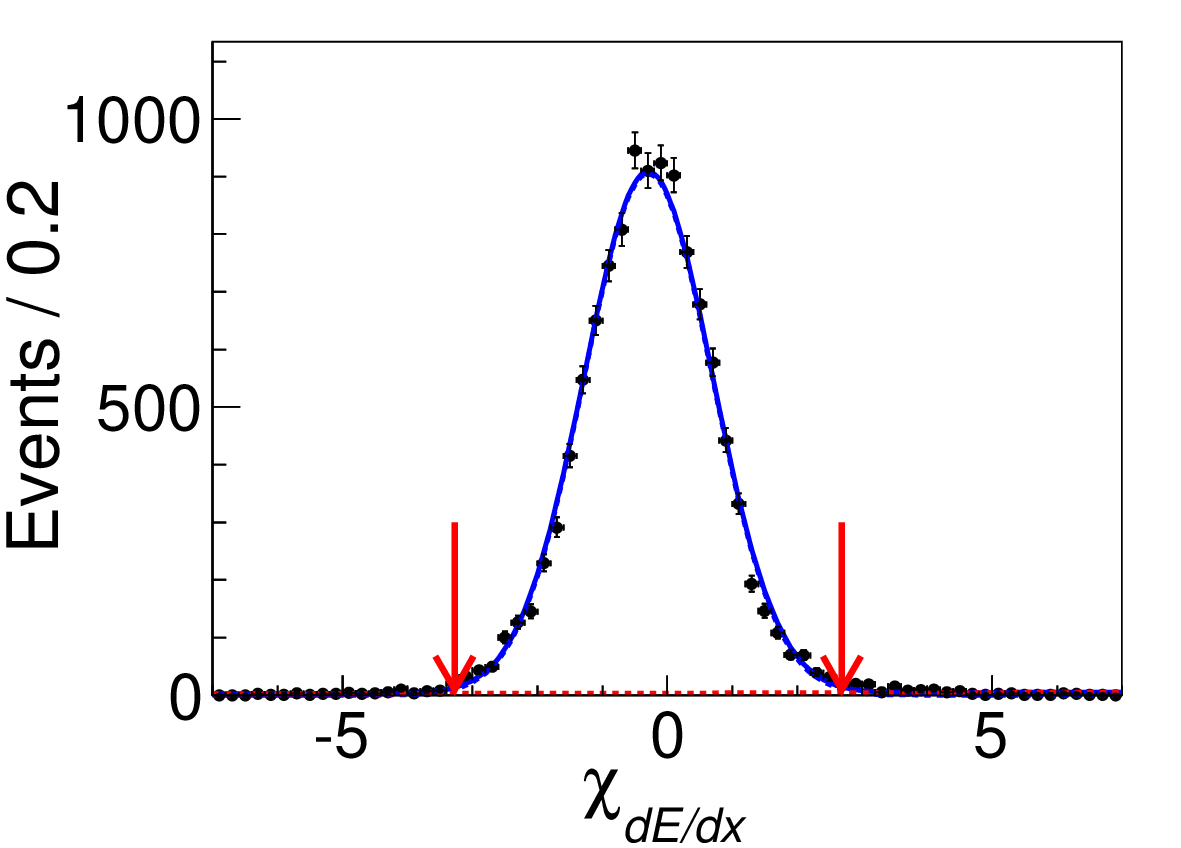}
\includegraphics[width=0.23\textwidth]{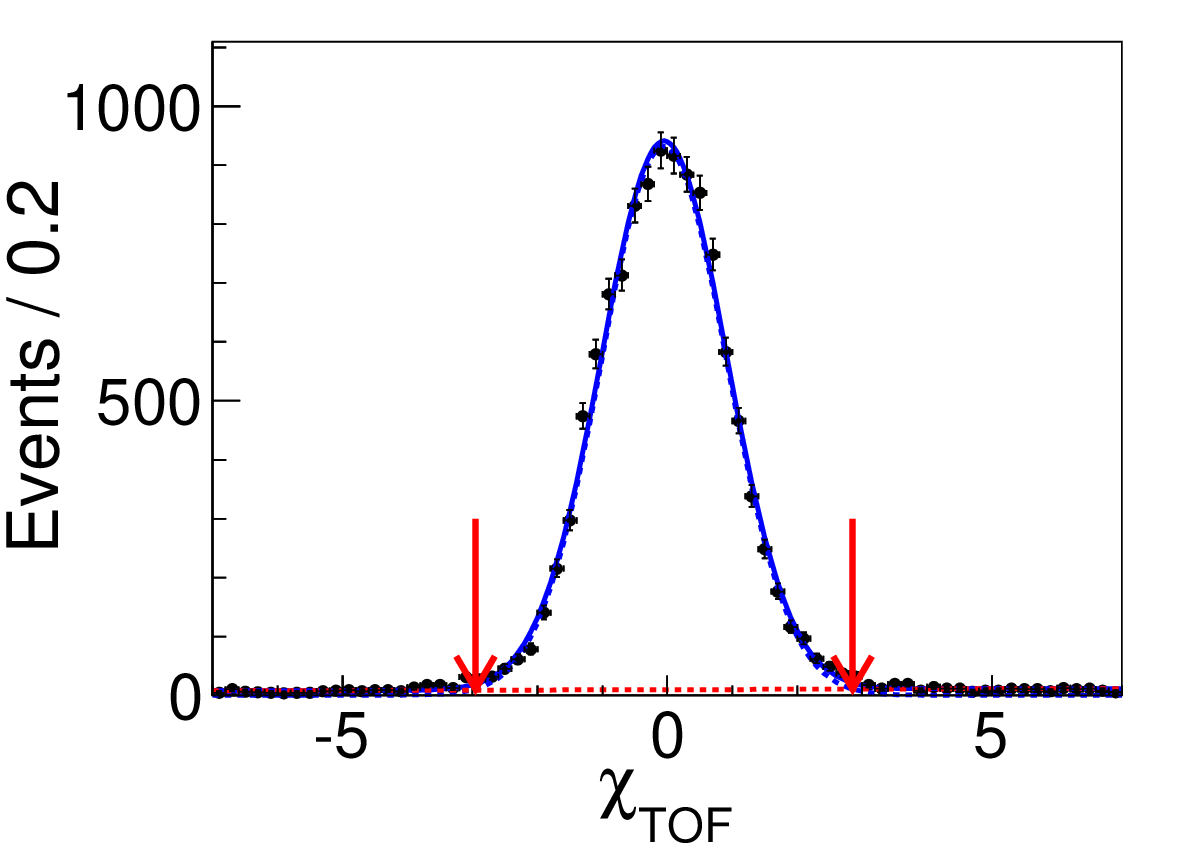}
\figcaption{Fits to $\chi_{dE/dx}$ (left) and $\chi_{\rm TOF}$ (right) of candidates in the momentum interval $0.52$-$0.57$ GeV/$c$ in the 2016 data sample. A Gaussian function marked with blue dashed line and a first-order Chebychev polynomial marked with red dashed line.}
\label{FIGESY}
\end{center}

The $d$ PID efficiencies of data and MC simulation as well as their differences in different momentum ranges and data taking periods are shown in Table \ref{TABSI}.  Figure~\ref{FIGESYY} shows the variation of the $d$ PID efficiency of data with momentum interval. We can see that the $d$ PID efficiencies of the data samples from $2011$ to $2017$ are higher than $97\%$, while those of the data samples from $2020$ to $2021$ are slightly low but still not less than $95\%$. These indicate that the $dE/dx$ method is an effective approach for the $d$ identification.
\begin{center}
\centering
\tiny
\tabcaption{\label{TABSI} The $d$ PID efficiencies of data and MC simulation as well as their differences in different momentum ranges and data taking periods.}
\begin{tabular}{c|c|c|c|c}
\hline
\hline
Year&$p$~(GeV/$c$)&$\varepsilon_{\rm data}(\%)$&$\varepsilon_{\rm MC}(\%)$&$\Delta_{\rm diff}\pm\Delta_{\rm fit}(\%)$\\
\hline
\multirow{4}{1.5em}{\makecell{$2011$}}&$0.52$-$0.57$&$99.63\pm1.79$&$100.00\pm0.03$&$0.37\pm0.01$\\
&$0.57$-$0.62$&$99.34\pm1.47$&$100.00\pm0.02$&$0.66\pm0.30$\\ 
&$0.62$-$0.67$&$99.24\pm1.58$&$100.00\pm0.02$&$0.76\pm0.47$\\
&$0.67$-$0.72$&$98.57\pm1.78$&$100.00\pm0.04$&$1.43\pm0.59$\\
\hline 
\multirow{4}{1.5em}{\makecell{$2013$}}&$0.52$-$0.57$&$97.62\pm1.07$&$100.00\pm0.02$&$2.38\pm0.43$\\ 
&$0.57$-$0.62$&$97.27\pm0.89$&$100.00\pm0.01$&$2.73\pm0.43$\\
&$0.62$-$0.67$&$98.28\pm0.95$&$100.00\pm0.01$&$1.72\pm0.36$\\ 
&$0.67$-$0.72$&$98.70\pm1.07$&$100.00\pm0.01$&$1.30\pm0.26$\\ 
\hline
\multirow{4}{1.5em}{\makecell{$2014$}}&$0.52$-$0.57$&$100.00\pm1.27$&$100.00\pm0.03$&$0.00\pm0.00$\\
&$0.57$-$0.62$&$99.43\pm1.04$&$100.00\pm0.01$&$0.57\pm0.15$\\
&$0.62$-$0.67$&$99.01\pm1.12$&$100.00\pm0.01$&$0.99\pm0.13$\\ 
&$0.67$-$0.72$&$100.00\pm1.25$&$100.00\pm0.01$&$0.00\pm0.00$\\ 
\hline
\multirow{4}{1.5em}{\makecell{$2016$}}&$0.52$-$0.57$&$99.94\pm0.94$&$100.00\pm0.02$&$0.06\pm0.39$\\ 
&$0.57$-$0.62$&$99.99\pm0.79$&$100.00\pm0.01$&$0.01\pm0.66$\\
&$0.62$-$0.67$&$99.75\pm0.87$&$99.95\pm0.01$&$0.20\pm0.72$\\ 
&$0.67$-$0.72$&$99.90\pm0.97$&$99.88\pm0.01$&$0.02\pm0.76$\\ 
\hline
\multirow{4}{1.5em}{\makecell{$2017$}}&$0.52$-$0.57$&$98.39\pm0.95$&$100.00\pm0.02$&$1.61\pm0.17$\\ 
&$0.57$-$0.62$&$98.11\pm0.79$&$99.47\pm0.01$&$1.37\pm0.40$\\
&$0.62$-$0.67$&$98.62\pm0.86$&$99.71\pm0.01$&$1.09\pm0.39$\\ 
&$0.67$-$0.72$&$99.21\pm0.96$&$98.86\pm0.01$&$0.35\pm0.30$\\ 
\hline
\multirow{4}{1.5em}{\makecell{$2020$}}&$0.52$-$0.57$&$95.07\pm1.04$&$100.00\pm0.02$&$4.93\pm0.95$\\ 
&$0.57$-$0.62$&$95.87\pm0.87$&$100.00\pm0.01$&$4.13\pm0.22$\\
&$0.62$-$0.67$&$96.47\pm0.94$&$100.00\pm0.01$&$3.53\pm0.27$\\ 
&$0.67$-$0.72$&$97.06\pm1.07$&$100.00\pm0.01$&$2.94\pm0.22$\\ 
\hline
\multirow{4}{1.5em}{\makecell{$2021$}}&$0.52$-$0.57$&$95.55\pm1.39$&$100.00\pm0.03$&$4.45\pm0.84$\\ 
&$0.57$-$0.62$&$96.14\pm1.16$&$100.00\pm0.02$&$3.86\pm0.74$\\
&$0.62$-$0.67$&$95.90\pm1.26$&$100.00\pm0.01$&$4.10\pm0.57$\\ 
&$0.67$-$0.72$&$97.52\pm1.40$&$100.00\pm0.01$&$2.48\pm0.48$\\ 
\hline
\hline
\end{tabular}
\end{center}

\begin{center}
\centering
\includegraphics[width=0.46\textwidth]{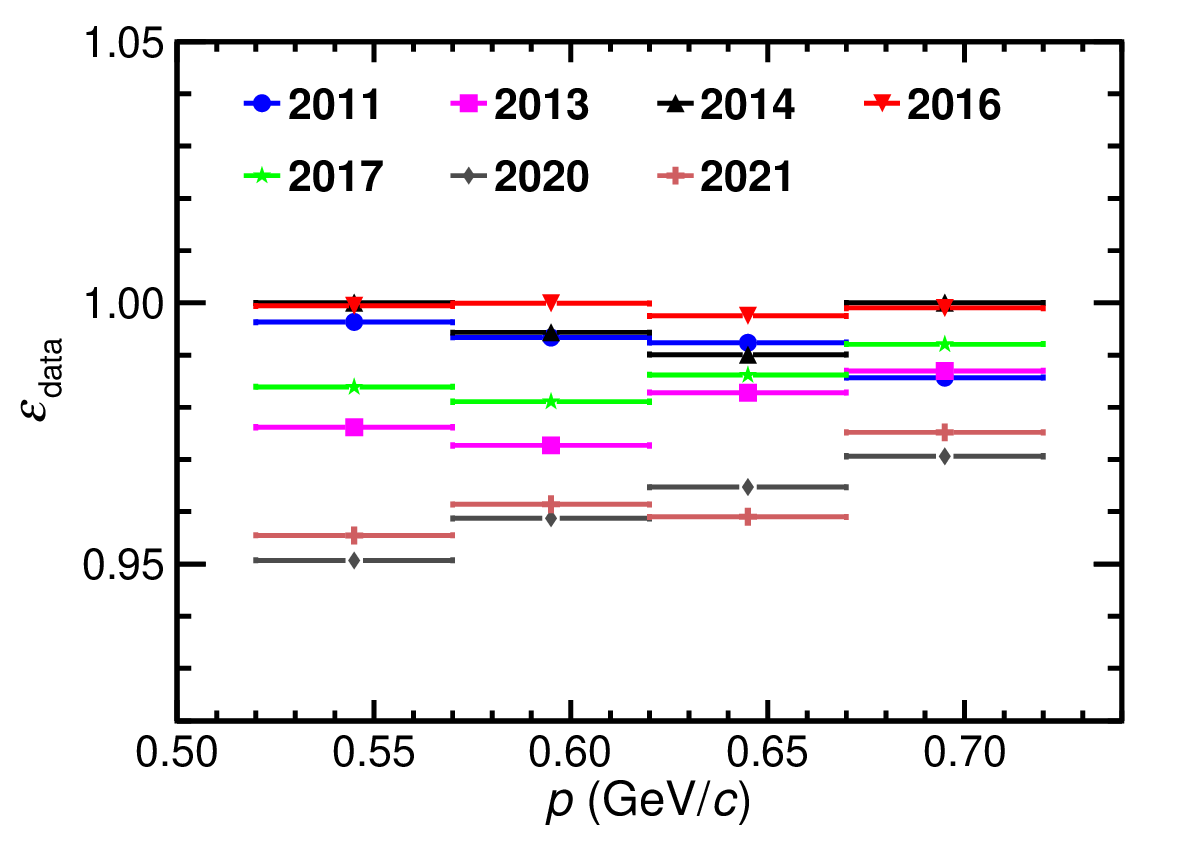}
\figcaption{Variation of the $d$ PID efficiency of data with momentum interval.}
\label{FIGESYY}
\end{center}

\subsection{Efficiency difference between data and MC simulation}
The difference of the $d$ PID efficiencies between data and MC simulation is determined by
\begin{equation}
\Delta_{\rm diff} = \lvert\dfrac{\varepsilon_{\rm data} - \varepsilon_{\rm MC}}{\varepsilon_{\rm MC}}\rvert,
\label{EQWU}
\end{equation}
where $\varepsilon_{\rm data}$ and $\varepsilon_{\rm MC}$ represent the efficiencies of data and MC simulation, respectively. The uncertainty of $\Delta_{\rm diff}$ is assigned as the difference between the fit and sideband methods and is calculated by
\begin{equation}
\Delta_{\rm fit} = \sqrt{\left(\frac{\Delta \varepsilon_{\rm data}}{\varepsilon_{\rm MC}}\right)^2 + \left(\frac{\varepsilon_{\rm data}\Delta \varepsilon_{\rm MC}}{\varepsilon_{\rm MC}^2}\right)^2},
\label{EQBA}
\end{equation}
\begin{equation}
\Delta {\varepsilon_{\rm data}} = \lvert \dfrac{\varepsilon_{\rm data}^{\rm fit} - \varepsilon_{\rm data}^{\rm side}}{\varepsilon_{\rm data}^{\rm fit}} \rvert,
\label{EQLIU}
\end{equation}
\begin{equation}
\Delta {\varepsilon_{\rm MC}} = \lvert \dfrac{\varepsilon_{\rm MC}^{\rm fit} - \varepsilon_{\rm MC}^{\rm side}}{\varepsilon_{\rm MC}^{\rm fit}} \rvert,
\label{EQQI}
\end{equation}
where $\varepsilon_{\rm data}^{\rm fit}$, $\varepsilon_{\rm data}^{\rm side}$, $\varepsilon_{\rm MC}^{\rm fit}$ and $\varepsilon_{\rm MC}^{\rm side}$ represent the efficiencies of data and MC simulation with the fit and sideband methods, respectively.

The obtained results are summarized in Table \ref{TABSI}. Good data-MC consistency in the $d$ PID efficiencies can be seen. 

\section{Summary and prospect}
In this paper, we report the study of the identification of $d$ in the momentum range of $0.52$-$0.72$ GeV/$c$, using $e^+e^-$ collision data taken in the c.m. energy range of $4.009$-$4.946$ GeV at BESIII. Based on the $dE/dx$ method, the $d$ PID efficiencies of data are higher than $95\%$, with a maximum difference of $(4.93\pm0.95)\%$ between data and MC simulation. For the data samples collected from 2011 to 2017, the $d$ PID efficiencies are higher than $97\%$. This shows good performance of the $d$ identification. Additional methods are expected to study the identification of $d$ in higher or lower momentum ranges. In addition, the data samples taken at higher c.m. energies by BESIII in the near future will offer new opportunity to further explore the $d$ identification~\cite{yy51,yy52}.

\section{Acknowledgments}
The authors are grateful to the BESIII software group for the profitable discussions. We express our gratitude to the BESIII Collaboration and the BEPCII team  for their strong support. This work is supported by the National Natural Science Foundation of China under Projects Nos. $11975118$, $12205141$ and $12375071$.
\end{multicols}

\vspace{-1mm}
\centerline{\rule{80mm}{0.1pt}}
\vspace{2mm}

\begin{multicols}{2}

\end{multicols}

\clearpage
\end{CJK*}
\end{document}